\documentclass[fleqn,usenatbib,useAMS,onecolumn]{mnras}
\usepackage{booktabs}
\usepackage{float}
\usepackage{array}
\usepackage{float}
\usepackage{graphicx}	% Including figure files
\usepackage{amsmath}	% Advanced maths commands
\usepackage{multicol}        % Multi-column entries in tables
\usepackage{bm}		% Bold maths symbols, including upright Greek
\usepackage{pdflscape}	% Landscape pages
\usepackage[T1]{fontenc}
\usepackage{ae,aecompl}

\usepackage{newtxtext,newtxmath}

\newcommand\pair{$e^-/e^+$\,}

\newcommand{\revise}{}
\title[Optical depth constraints on GRB 221009A line]{Robust Constraints on the Physics of the MeV Emission Line in GRB 221009A from Optical Depth Arguments}

\author[S. Yi et al.]{Shu-Xu Yi$^{1}$\thanks{Contact e-mail: \href{sxyi@ihep.ac.cn}{sxyi@ihep.ac.cn}}, Zhen Zhang$^1$\thanks{zhangzhen@ihep.ac.cn}, Emre Seyit Yorgancioglu$^{1}$, Shuang-Nan Zhang$^{1}$, Shao-Lin Xiong$^{1}$, Yan-Qiu Zhang$^{1}$%
\\%
$^{1}$Key Laboratory of Particle Astrophysics, Institute of High Energy Physics, Chinese Academy of Sciences, Beijing 100049, China}
\begin{document}
\maketitle
\begin{abstract}
The brightest-of-all-time gamma-ray burst (GRB), GRB 221009A, is the first GRB observed to have emission line (up to 37 MeV) in its prompt emission spectra. It is naturally explained as \pair annihilation line that was Doppler boosted in the relativistic jet of the GRB.  In this work, we repeatedly apply the simple optical depth argument to different physical processes necessary to produce an observable \pair annihilation line. This approach results in robust constraints on the physics of the line: We conclude that in GRB 221009A, the \pair pairs were produced at a radius greater than $4.3\times 10^{15}$\,cm from the central engine, and annihilated in a region between $1.4\times 10^{16}$\,cm and $4.3\times 10^{16}$\,cm. From these constraints, we established a self-consistent picture of \pair production, cooling, and annihilation. We also derived a criterion for pair production in the GRB prompt emission: $E_{\rm{iso}} \gtrsim3.3\times 10^{53} E_{\rm{peak},100} (1+z) R^2_{\rm{prod},16}~\text{erg}$. Using this criterion, we find tens of candidate GRBs that could have produced \pair in prompt emissions to annihilate. GRB 221009A is with the highest likelihood according to this criterion. We also predict the presence of a thermal radiation, with a time-evolving black body temperature, sweeping through soft X-ray during the prompt emission phase.
%The brightest-of-all-time gamma-ray burst (GRB), GRB 221009A, is the first GRB observed to have emission line (up to 37 MeV) in its prompt emission spectra. The central energy of the line follows a power-law function of time with an index of about -1, and its flux evolves with a power-law index of about -2. It is naturally explained as \pair annihilation line that was Doppler boosted in the highly relativistic jet of the GRB.  In this work, we repeatedly apply the simple optical depth argument to different physical processes necessary to produce an observable \pair annihilation line. This approach results in robust constraints on the physics of the line: We conclude that in GRB 221009A, the \pair pairs were produced at a distance greater than $4.3\times 10^{15}$\,cm from the central engine, and annihilated in a region between $1.4\times 10^{16}$\,cm and $4.3\times 10^{16}$\,cm. From these constraints, we established a self-consistent picture of \pair production, cooling, and annihilation. We also derived a criterion for pair production in the GRB prompt emission: $E_{\rm{iso}} \gtrsim3.3\times 10^{53} E_{\rm{peak},100} (1+z) R^2_{\rm{prod},16}~\text{erg}$. Using this criterion, we find  tens of candidate GRBs that could have produced \pair in prompt emissions to further annihilate. GRB 221009A is with the highest likelihood according to this criterion. We also predict the presence of a thermal radiation of GRB 221009A, with a time-evolving black body temperature, sweeping through soft X-ray during the prompt emission phase. 
\end{abstract}


\section{Introduction}\label{sec:intro}
GRB 221009A was observed on the 9th October 2022 with many telescopes, such as {\it Fermi} \citep{Fermi-trigger,FermiGBM}, {\it Insight}-HXMT \citep{Insight}, GECAM-C \citep{An23}, {\it Swift} \citep{Swift}, Konus-WIND \citep{Konus}, {\it AGILE} \citep{Agile}, {\it INTEGRAL} \citep{Integral}, GRANDMA \citep{grandma} and LHAASO \citep{Lhaaso}. %Due to its extraordinary brightness, many detectors suffered from instrumental problems such as data saturation that bias their observation. The dedicated design in detector and read-out electronics of GECAM-C, along with its special operational mode during the burst, allowed it to obtain unbiased data even in the brightest phase of the burst. The unsaturated and high-resolution GECAM-C data enabled \cite{An23} to accurately study the spectral and temporal features of this burst and provided a reliable measurement of its record-breaking isotropic equivalent energy $E_{\rm{iso}}$ as high as $1.5\times10^{55}\,$erg. 
{\revise Interestingly, the isotropic equivalent energy $E_{\rm{iso}}$ of the burst is $1.5\times10^{55}\,$erg \citep{An23}, setting a new record.} 
{\revise This burst} was thus identified as the brightest-of-all-time (BOAT) by its peak flux, fluence and $E_{\rm{iso}}$, whose event rate was estimated to be once-in-10,000-year \citep{Burns23} {(\revise although \citealt{2023arXiv230207891M} came to a slightly higher event rate.)}. Additionally, from the observation of the jet break in its early afterglow detected both in keV-MeV by {\it Insight}-HXMT, GECAM-C and Fermi/GBM \citep{An23,Zhengchao24} and in TeV by LHAASO \citep{2023Sci...380.1390L}, the opening angle of the jet was estimated to be $\sim0.7^\circ$, making it one of the most collimated GRBs ever recorded, {\revise despite the debate surrounding the early identification of the jet break \citep{2023ApJ...946L..23L,2023SciA....9I1405O}.}

%\bibitem[Laskar et al.(2023)]{2023ApJ...946L..23L} Laskar, T., Alexander, K. D, Margutti, R., et al.\ 2015, \apjl, 946, L23. 10.3847/2041-8213/acbfad
%\bibitem[O'Connor et al.(2023)]{2023SciA....9I1405O}  O'Connor, B., Troja, E., Ryan, G., et al.\ 2023, Science Advances, , vol. 9, issue 23, id. eadi1405, doi: 10.1126/sciadv.adi1405 %title:A structured jet explains the extreme GRB 221009A

The uniqueness of this burst lies beyond its brightness and narrow jet beam. 
%(Shortly after the discovery of the burst,) 
\cite{Ravasio23} reported emission line structures in its spectra from 280 s to 360 s after the trigger time in the data of {\it Fermi}/GBM. The central energy of the emission line was found to be time-varying from about 12 MeV to 6 MeV. Meanwhile, \cite{ZYQ24} jointly analyzed the data from {\it Fermi}/GBM and GECAM-C and found emission line structures varying from 37 MeV to 6 MeV from 246 s to 360 s after trigger time. More importantly, thanks to the larger time span of their detection of the emission line, \cite{ZYQ24} found the central energy of the line evolves as a power-law function with time, with a power-law index of approximately -1, and the flux of the line also evolved as a power-law function with time, with a power-law index of approximately -2. These power-law time-evolution features provide critical evidence that the apparent line structures originate from this GRB rather than from possible instrumental effects or background. 

A natural explanation of the observed line is that the MeV line originated from \pair annihilation in the relativistic jet, which was Doppler boosted to different observed central energies at different times (see \citealt{Ravasio23,ZYQ24,ZZ24}; however, \cite{Wei24} proposed a novel model explaining the MeV line as relativistic hydrogen-like heavy ions atomic line). \cite{ZZ24} (Zhang24 hereafter) conducted a detailed study based on the \pair annihilation framework, where they attributed the time evolution of the line central energy to emission from increasing higher latitude of the jet. According to this assumption, they were able to infer the Lorentz factor and radius of the jet prompt emission region. Zhang24 also essentially excluded some of the alternative mechanisms such as nuclear de-excitation of some radioisotope synthesized during the burst process. Most recently, \cite{2024arXiv240716241P} analyses the details of the balanced process $\gamma+\gamma\leftrightarrow e^++e^-$ in order to produce the observed MeV line and deduced constraints on the physics of the source and explained the rarity of such line in GRB. 

In this work, we are going to demonstrate that, only if we assume the \pair annihilation picture, can we apply the simple optical depth equation repeatedly, namely $$\tau=nl\sigma=n^{\prime}l^{\prime}\sigma,$$ 
%(with different particle number density and cross sections in different physical processes).
where $\tau$, $n$, $l$ and $\sigma$ are the optical depth, number density of particles, geometrical depth and cross section of the corresponding physical process, respectively. The quantities with (without) primes represent those in the comoving (rest) frame, while the cross-section is a Lorentz-invariant. As we will show in the next section, by requiring $\tau>1$ or $\tau<1$ for different physical processes, we can obtain many tight constraints on the physical origin of the lines. Such constraints include the positions of the sites of \pair annihilation and production. We can see that these constraints are consistent with Zhang24; however, our results are independent of the assumption of high-latitude effects placed in Zhang24 and also independent of the value of the bulk Lorentz factor and the jet opening angle. Based on these inferences, we provide a self-consistent picture from \pair production to annihilation. Furthermore, we point out the minimum required $E_{\rm{iso}}$ needed to produce \pair in prompt emission as function of its $E_{\rm{peak}}(1+z)$. Our conclusion is that, among tens of other candidates, GRB 221009A has the highest likelihood that there were \pair produced in the prompt emission via $\gamma\gamma$ process. In the following section, after we summary our main conclusion on the nature of the line, we discuss on the characteristic of the predicted relic thermal radiation from the equilibrium state of \pair production. 

\section{Implication on the physical origin of the emission line}

\subsection{The condition of emergence of the line emission} 
An emission line can only be observed if the optical depth of the emission region is less than unity. The optical depth caused by the scattering from electrons/positrons are: 
\begin{equation}
\label{eq:tau}
%\tau=2\sigma_{\rm T}n^\prime_\pm\delta R_{\rm{line}}\Gamma,
\tau=2\sigma_{\rm T}n^\prime_\pm\delta R^{\prime}_{\rm{line}},
\end{equation}
where $\sigma_{\rm T}$ is the Thompson scattering cross section,  $n^\prime_\pm$ is the number density of $e^-/e^+$ pairs in the comoving frame, and $\delta R^{\prime}_{\rm{line}}$ is the geometrical width of the line emitting region in the comoving frame. Note that this equation for optical depth is generally valid in GRB 221009A, as the time scale being considered is much shorter than the dynamical time scale \citep{ZZ24,2024arXiv240716241P}, required by the narrow line width of $\sim10\%$, ensuring that the region does not significantly expand.
%{\red We consider this as a good approximation} in our following arguments, and this point is discussed in details in Zhang24 and \cite{ZZ24,2024arXiv240716241P}.
%which is in the same order of magnitude of the distance from this region to the central engine, 
%and $\Gamma$ is the bulk Lorentz factor of the jet. 
%The latter two terms composed the geometrical size of the emitting region in the comoving frame.  
The number density $n^\prime_\pm$ is the ratio between the pair number $N_\pm$ and the comoving volume of the emitting region $V^\prime$; the former can be estimated from the observed total energy in the line emission. The total line emission energy in the comoving frame can be calculated as:
\begin{equation}
E^\prime_{\rm{line}}=N_\pm m_ec^2=4\pi D^2_{\rm{L}}f_b(1+z)\int^{t_f}_{t_i}\frac{\mathcal{F}(t)}{\mathcal{D}(t)}dt,
\end{equation}
where $D_{\rm{L}}$ is the luminosity distance of the GRB, $f_b$ is the beaming factor of the jet, $\mathcal{F}(t)$ is the observed line flux at instant $t$ and $\mathcal{D}(t)$ is the Doppler factor as function of observer's time. The integral is conducted from the initial time $t_i$ to the final time $t_f$ when the line appears. The Doppler factor equals to the ratio between the observed central line energy $\mathcal{E}_{\rm{line}}$ and $m_ec^2=511$\,keV.  From the observation \citep{ZYQ24,ZZ24}, we know that the observed $\mathcal{F}$ and $\mathcal{E}_{\rm{line}}$ evolve with time in power laws, with the reference time $t_0=226$\,s, the power-law indices -2 and -1,  and the normalization factors $\mathcal{F}_0=0.02$\,erg/cm$^2$/s and $\mathcal{E}_{\rm{line},0}=8.4\times10^5$\,keV respectively. Therefore,  the above equations can be worked out as:
\begin{equation}
N_\pm=4\pi D^2_{\rm{L}}f_b(1+z)\frac{\mathcal{F}_0}{\mathcal{E}_{{\rm{line},0}}}\ln\frac{t_f-t_0}{t_i-t_0}.
\label{eq:3}
\end{equation}
The comoving volume of the emitting region can be estimated as:
\begin{equation}
%V^\prime=4\pi f_b R_{\rm{line}}^2\delta R_{\rm{line}}\Gamma.
V^\prime=4\pi f_b R_{\rm{line}}^2\delta R^{\prime}_{\rm{line}}.
\label{eq:4}
\end{equation}
Taking the expressions of $N_\pm$ and $V^\prime$ into equation (\ref{eq:tau}), we find
\begin{equation}
\tau=2\sigma_{\rm T}D^2_{\rm{L}}(1+z)\frac{\mathcal{F}_0}{\mathcal{E}_{\rm{line},0}}\ln\frac{t_f-t_0}{t_i-t_0}R_{\rm{line}}^{-2}.
\label{eq:tau_expression}
\end{equation}
The requirement that $\tau<1$ immediately gives a lower limit of the line emission radius:
\begin{equation}
R_{\rm{line}}>\sqrt{2\sigma_{\rm T}D^2_{\rm{L}}(1+z)\frac{\mathcal{F}_0}{\mathcal{E}_{\rm{line},0}}\ln\frac{t_f-t_0}{t_i-t_0}}=1.4\times10^{16}\,\text{cm}.
\label{eq:6}
\end{equation}
We take the values $t_i=250$\,s, $t_f=350$\,s \citep{ZYQ24}, $D_{\rm{L}}=745$\,Mpc and $z=0.151$ \citep{2023arXiv230207891M} in the above calculation. Note that the actual line occurrence duration can be larger than that was observed. Therefore, equation \ref{eq:6} serves as a conservative lower limit.

\subsection{The condition of pair annihilation}
The annihilation cross section for a pair of free $e^-/e^+$ in the non-relativistic regime is:
\begin{equation}
\sigma_{\rm{anni}}=\frac{3}{8}\sigma_{\rm T}/\beta_{\rm{rel}},
\end{equation}
where $\beta_{\rm{rel}}$ is the relative velocity between the pair (divided by $c$). The directional averaged $\beta_{\rm{rel}}\sim\beta^\prime_e$, where the latter is the typical velocity of \pair in the comoving frame (see Appendix). The effective ``optical" depth of \pair annihilation can be defined as:
\begin{equation}
%\tau_{\rm{anni}}=\sigma_{\rm{anni}}n^\prime_\pm\delta R_{\rm{line}}\Gamma,
\tau_{\rm{anni}}=\sigma_{\rm{anni}}n^\prime_\pm\delta R^{\prime}_{\rm{line}},
\end{equation}
where $\tau_{\rm{anni}}>1$ is the region where the \pair can annihilate efficiently. Compared with equation (\ref{eq:tau}), one can find that
\begin{equation}
\tau_{\rm{anni}}=\frac{3}{16\beta_{\rm{rel}}}\tau. 
\label{eq:tau_tau_anni}
\end{equation}
The observation of the line emission indicates that $\tau_{\rm{anni}}>1$ and $\tau<1$ at the same time, which in return requires $\beta<3/16\sim0.18$. This is in agreement with Zhang24's statement that, the electrons need to be cooled down to non-relativistic $\beta^\prime_e\sim0.1$ in order to annihilate efficiently. In Zhang24, the authors found that the cooling from synchrotron radiationn and inverse Compton scattering are sufficient to cool down the \pair to the required velocity in the comoving frame. However, there lack effective mechanisms to further cool the electrons down to $\beta^\prime_e<0.01$ in the jet environment, which corresponds to the kinetic energy of tens of electron Volts. 
%{\red In fact, the ratio of line width to central energy is much larger than 0.01 in observations, suggesting that $\beta^\prime_e>0.01$.}
We take the expression of $\tau$ in equation (\ref{eq:tau_expression}) into equation (\ref{eq:tau_tau_anni}) to find:
\begin{equation}
\tau_{\rm{anni}}=\frac{3}{8\beta_{\rm{rel}}}\sigma_{\rm T}D^2_{\rm{L}}(1+z)\frac{\mathcal{F}_0}{\mathcal{E}_{\rm{line},0}}\ln\frac{t_f-t_0}{t_i-t_0}R_{\rm{line}}^{-2}.
\label{eq:10}
\end{equation}
In the above equation, we insert the condition that $\tau_{\rm{anni}}>1$ and $\beta_{\rm{rel}}>0.01$ to obtain another limit on $R_{\rm{line}}$: 
\begin{equation}
R_{\rm{line}}<\sqrt{\frac{3}{8}\times100\sigma_{\rm T}D^2_{\rm{L}}(1+z)\frac{\mathcal{F}_0}{\mathcal{E}_{\rm{line},0}}\ln\frac{t_f-t_0}{t_i-t_0}}=4.3\times10^{16}\,\text{cm}.
\label{eq:11}
\end{equation}
Therefore, we see from the simple arguments on the condition of optical depth, we can already set a quite stringent constraint on the location of the \pair annihilation to be between $1.4\times10^{16}$\,cm and $4.3\times10^{16}$\,cm, which is in agreement with the conclusion of Zhang24. However, our conclusion does not depend on the value of the bulk Lorentz factor and the opening angle of the jet. Note that the exact numbers for the upper and lower limits should take into account the observation uncertainties of all the quantities in equations (\ref{eq:6}) and (\ref{eq:11}), which is not done here for the numbers will not be significantly changed.

\subsection{The origin of $e^+/e^-$ pairs}
\begin{figure*}
\centering
\includegraphics[width=0.8\textwidth]{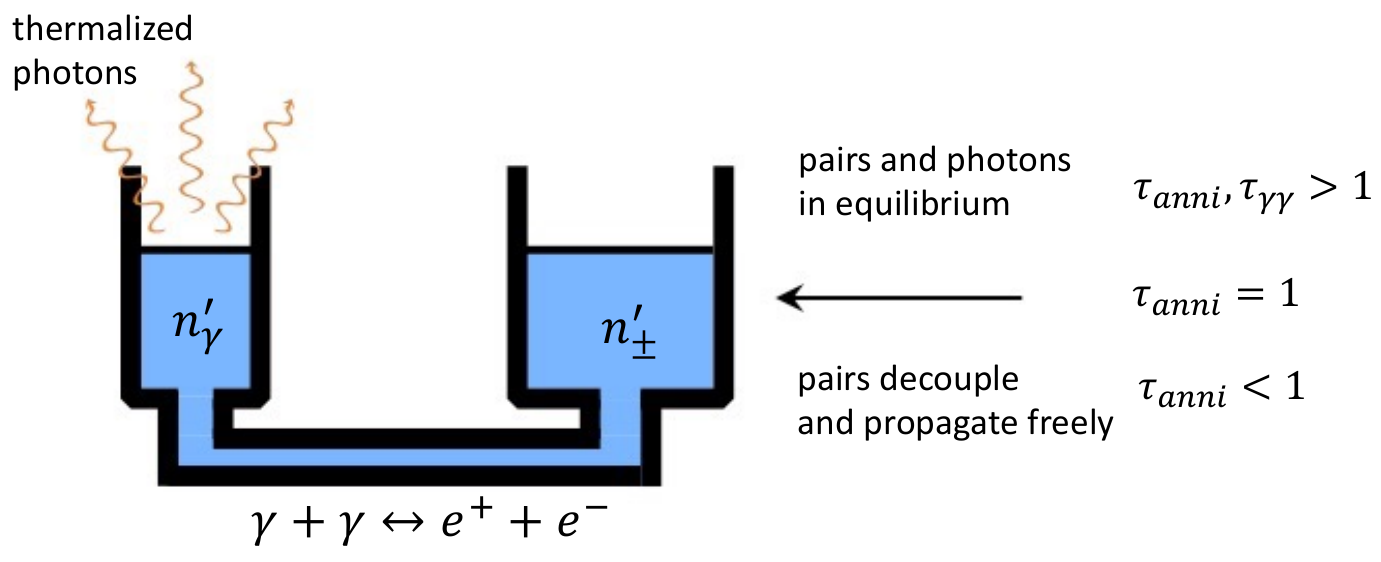}
\caption{In this cartoon illustration, we draw an analogy between the equilibrium process of \pair production/annihilation and a hydrostatic equilibrium system with two connected beakers. The liquid volume on the left represents the number density of high-energy photons involved in pair production $n^\prime_\gamma$, while that on the right represents $n^\prime_\pm$. In such a hydrostatic equilibrium system, the liquid levels on both sides remain consistent, analogous to $n^\prime_\gamma$  and $n^\prime_\pm$ tracking each other in equilibrium. The reduction of the liquid on the left due to evaporation analogous to the decrease in high-energy photons due to thermalization. During this process, the liquid representing $n^\prime_\pm$ on the right also decreases. When the liquid level drops below the mark labeled $\tau_{\rm{anni}} = 1$, the electron-positron pairs decouple from the equilibrium state \citep{Ruffini:1999ta,Ruffini:2000yu,Bianco:2001fw} and begin to propagate freely.}
\label{fig:1a}
\end{figure*}

Denote the radius where the \pair were produced and started to transmit freely as $R_{\rm{prod}}$. Then $\tau_{\rm{anni}}$ should be less than unity in the range from $R_{\rm{prod}}$ to $R_{\rm{line}}$, in order that the pairs would not be exhausted before $R_{\rm{line}}$. Therefore, similar to equation (\ref{eq:10}), we have:
\begin{equation}
\tau_{\rm{anni}}=\frac{3}{8\beta_{\rm{rel},0}}\sigma_{\rm T}D^2_{\rm{L}}(1+z)\frac{\mathcal{F}_0}{\mathcal{E}_{\rm{line},0}}\ln\frac{t_f-t_0}{t_i-t_0}R_{\rm{prod}}^{-2}<1,
\end{equation}
where $\beta_{\rm{rel},0}$ is the relative velocity between \pair  before the fast cooling. We thus have
\begin{eqnarray}
R_{\rm{prod}}&>&\sqrt{\frac{3}{8\beta_{\rm{rel},0}}\times\sigma_{\rm T}D^2_{\rm{L}}(1+z)\frac{\mathcal{F}_0}{\mathcal{E}_{\rm{line},0}}\ln\frac{t_f-t_0}{t_i-t_0}}\\&>&\sqrt{\frac{3}{8}\sigma_{\rm T}D^2_{\rm{L}}(1+z)\frac{\mathcal{F}_0}{\mathcal{E}_{\rm{line},0}}\ln\frac{t_f-t_0}{t_i-t_0}}\\
&=&4.3\times10^{15}\,\rm{cm}.
\end{eqnarray}
This indicates that the pairs were produced from a region far from the centre engine and well above the photosphere ($\sim10^{13}$\,cm). In Zhang24, \pair  were assumed to be produced by $\gamma\gamma$ process in the prompt emission region. This assumption is again consistent with our implication on $R_{\rm{prod}}$.  If $N_{\gamma,0}$ is the number of photons in the initial prompt emission, and a fraction of them $\eta$ participate in the $\gamma\gamma$ process. Then the number density of those photons is:
\begin{eqnarray}
n^\prime_\gamma&=&\eta N_{\gamma,0}/V^\prime_{\rm{prod}}\nonumber\\
%&=&\frac{\eta N_{\gamma,0}}{f_bR^2_{\rm{prod}}\delta R_{\rm{prod}}\Gamma}.
&=&\frac{\eta N_{\gamma,0}}{f_bR^2_{\rm{prod}}\delta R^{\prime}_{\rm{prod}}}.
\end{eqnarray}
The optical depth of $\gamma\gamma$ pair production is:
\begin{eqnarray}
%\tau_{\gamma\gamma}&=&\sigma_{\gamma\gamma}n^\prime_{\gamma}\delta R_{\rm{prod}}\Gamma\nonumber\\
\tau_{\gamma\gamma}&=&\sigma_{\gamma\gamma}n^\prime_{\gamma}\delta R^{\prime}_{\rm{prod}}\nonumber\\
&=&\sigma_{\gamma\gamma}\eta N_{\gamma,0}f_b^{-1}R^{-2}_{\rm{prod}},
\label{eq:17}
\end{eqnarray}
which is required to be larger than unity. The fraction $\eta$ is determined by the initial photon spectrum 
$n^\prime_\gamma(\nu^\prime)$. We emphasize here that, the correspondence between the initial spectrum and the observed spectrum can be complicated since the pair production process altered the initial spectrum. A number of photons $\eta N_{\gamma,0}$ will be converted into \pair pairs, and the number of these initially produced pairs equals $N_{\pm, 0}=\eta N_{\gamma,0}$. In the centre-of-momentum frame of the pair, the cross section $\sigma _{\rm{anni}}$ can be written as \citep{zhang19}:
\begin{equation}
\sigma _{\rm{anni}}=\frac{3}{8}\sigma_{\rm T}\frac{1}{\gamma^{\prime}_e +1} \left[\frac{\left(\gamma^{\prime\,2}_e+4 \gamma^{\prime}_e +1\right) \text{Log} \left(\sqrt{\gamma^{\prime\,2}_e-1}+\gamma^{\prime}_e \right)}{\gamma^{\prime\,2}_e-1}-\frac{\gamma^{\prime}_e +3}{\sqrt{\gamma^{\prime\,2}_e-1}}\right],
\end{equation}
where $\gamma^{\prime}_e=(1-\beta^{2}_{\rm rel})^{-1/2}\simeq(1-\beta^{\prime 2}_{\rm e})^{-1/2}$.
Correspondingly the cross section of the inverse process,
\begin{equation}
\sigma _{\gamma \gamma }=\frac{3}{8}\sigma_{\rm T}\gamma^{\prime\,2}_e \left[-\left(-\frac{1}{\gamma^{\prime\,4}_e}+\frac{2}{\gamma^{\prime\,2}_e}+2\right)\text{Log}\left(\left|\gamma^{\prime}_e -\sqrt{\gamma^{\prime\,2}_e-1}\right|\right)+\sqrt{1-\frac{1}{\gamma^{\prime\,2}_e}} \left(\frac{1}{\gamma^{\prime\,2}_e}+1\right)\right],
\end{equation}
where $\gamma^{\prime}_e\equiv h\nu^\prime/m_{e}c^2$. In the relativistic limit of $\beta^\prime_e\sim1$, $\sigma_{\rm{anni}}/\sigma_{\gamma\gamma}\sim0.4\gamma^{\prime}_e$, while in the non-relativistic limit of $\beta^\prime_e\ll1$, $\sigma_{\rm{anni}}/\sigma_{\gamma\gamma}\sim\beta^{\prime -2}_e$. Roughly, $\sigma_{\rm{anni}}/\sigma_{\gamma\gamma}\sim{\rm Max}\left(\beta^{\prime -2}_e,0.4\gamma^{\prime}_e\right)$. Consequently, the cross section $\sigma_{\gamma\gamma}$ of $\gamma\gamma\rightarrow e^+/e^-$ tends to be less than that of the inverse progress, and the produced pairs will immediately undergo the annihilation process. 
%{\blu However,  since the $\gamma\gamma\rightarrow e^+/e^-$ cross section $\sigma_{\gamma\gamma}$ is always less than the inverse progress, the produced \pair will immediately undergo the annihilation process. } 
Therefore, there should be a balance process between the pair production and annihilation, until the optical depth of interaction becomes less than unity \citep{Ruffini:1999ta,Ruffini:2000yu,Bianco:2001fw}. In the state of equilibrium, according to the Boltzmann equation, we expect: 
\begin{equation}
n^{\prime\, 2}_\gamma\sigma_{\gamma\gamma}\sim 2\,n^{\prime\, 2}_\pm\sigma_{\rm{anni}}\beta^\prime_e ~~~\Longleftrightarrow~~~ n^{\prime}_\gamma\sim n^{\prime}_\pm~\sqrt{2\,{\rm Max}\left(0.4\gamma^{\prime}_e,\frac{1}{\beta^{\prime}_e}\right)}. \label{eq:21}
%\simeq2\, \frac{3}{8} n^{\prime\, 2}_\pm\sigma_{\rm T}}
\end{equation}
In the above equation, we denote the number density of photons which satisfy the condition of pair production as $n^\prime_\gamma$. The number density of \pair $n^\prime_\pm$ thus tracks $n^\prime_\gamma$ (see also \citealt{2024arXiv240716241P}).  
%Nevertheless, this fraction of photons were in the equilibrium state and cooled (thermalized?) to lower energies with electrons and positrons eventually and finally cease to produce \pair. 
Nevertheless, this fraction of photons was in equilibrium with electrons and positrons and eventually cooled to lower energies 
 and finally the pair production ceases. Consequently, $n^\prime_\pm$ and $n^\prime_\gamma$ decrease until the pair production and annihilation process stops and decouple. We illustrate the above process in figure \ref{fig:1a}. 
 
\begin{figure}
\centering
\includegraphics[width=8 cm]{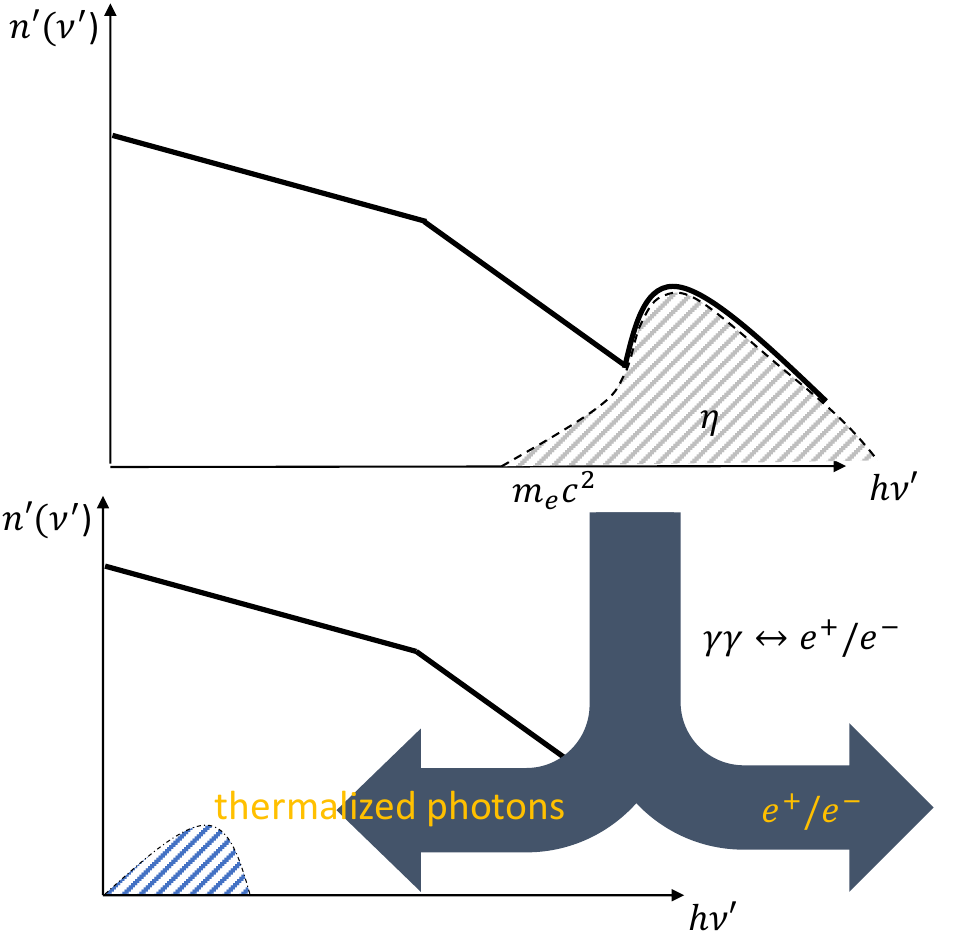}
\caption{The illustration of the pair production process. The initial spectrum of prompt radiation in the comoving frame is illustrated in the upper panel. A fraction $\eta$ of these photons underwent the pair production/annihilation process, so that roughly a half of these photons are converted into \pair, while the rest of them are thermalized and contributing to a thermal component in the observed prompt emission spectrum in the lower panel.}
\label{fig:1}
\end{figure}

 After this process, the initial photon spectrum is altered: its high energy end is converted into \pair and the thermalized photons appear at much lower energies. In figure \ref{fig:1}, we present the illustration of this process: The initial spectrum of the prompt radiation in the comoving frame is illustrated in the upper panel. A fraction $\eta$ of these photons underwent the pair production/annihilation process, so that roughly a half of these photons are converted into \pair, while the rest of them are thermalized and contribute to a thermal radiation in the observed prompt emission spectrum in the lower panel. We will discuss the observational prospect of the thermal radiation in the next section.   %Subsequently, the produced \pair decouple from these photons. Actually, in the equilibrium state, the plasma fluid, consisting of coupled \pair and photons, undergoes adiabatic expansion without significantly emitting radiative energy until they decouple, leading to the decrease in their effective temperature. At decoupling, the pairs and photons are in equilibrium at the same effective temperature, $T^{\prime}$, resulting in a thermalized component after decoupling.

Given an initial photon spectrum from detailed simulation, (as that illustrated in the upper panel of figure \ref{fig:1}, where the second bump at high energy can arise from the Self-Synchrotron Comptonization (SSC)), the fraction of photons which satisfy the pair production condition can be found with a Monte Carlo simulation, together with their total energy. The resulted number of \pair, the temperature of thermalized photons and the resulted prompt emission spectrum can thus calculated. A fitting with the simulated spectrum and the observed one can be used to set constraints on the detailed physics conditions. Such a study is beyond the scope of this paper, and will be conducted in a follow up work. 

\subsection{The lower limit of $E_{\rm{iso}}$ for sufficient pair production}

\begin{figure}
\centering
\includegraphics[width=12 cm]{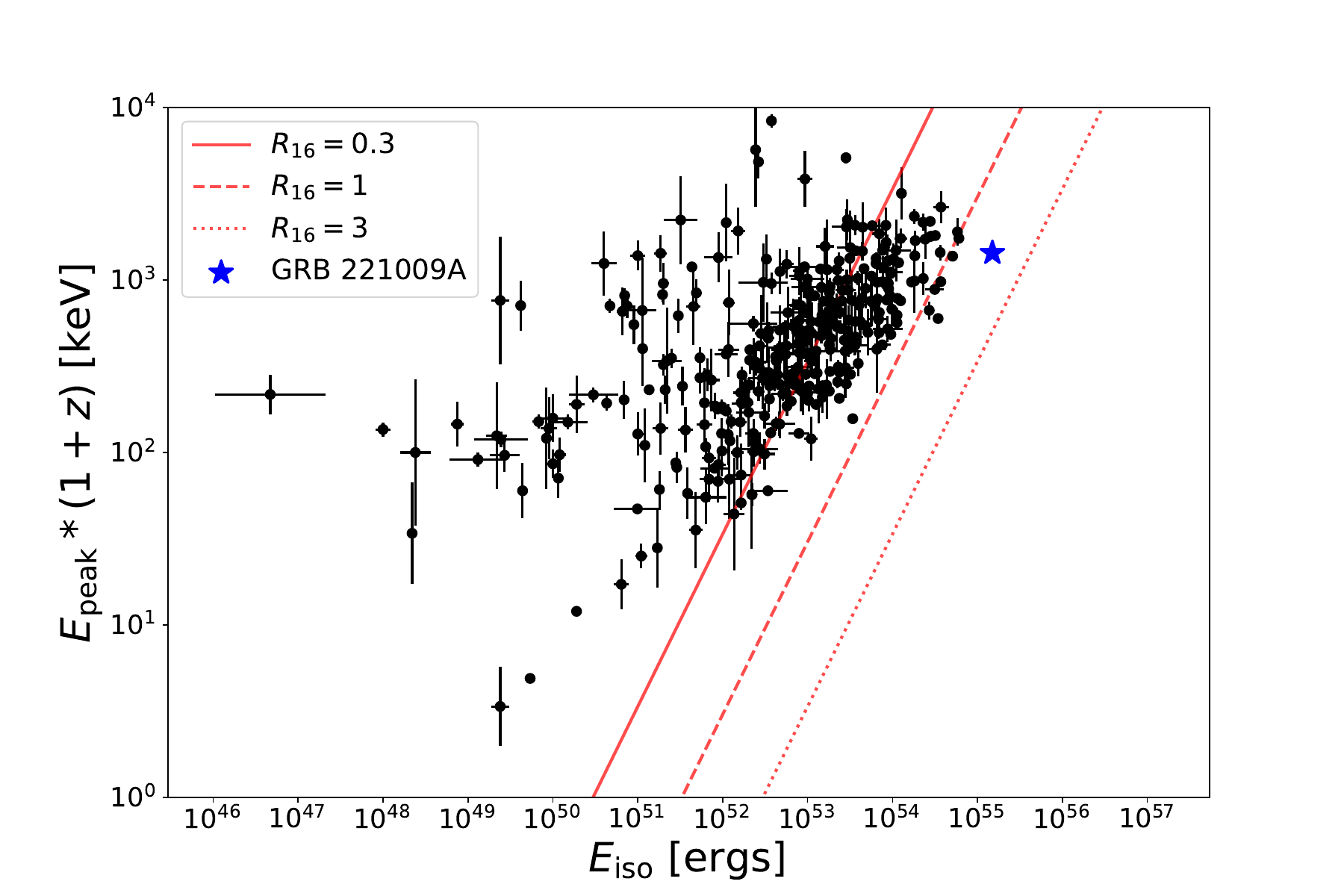}
\caption{The dashed line represents the $E_{\rm{iso}}$-$E_{\rm{peak}}(1+z)$ parameter space constraints  set by equation (\ref{eq:limit}), with $R_{\rm{prod, 16}}$ =0.3, 1, 3.  The data are taken from \citealt{An23}.}
\label{fig:limit}
\end{figure}

The rest of initial photons which do not participate the pair production will largely contribute to the observed prompt emission.  
The number of photons in the prompt emission can be approximated with $E^\prime_{\gamma, \rm{prompt}}/E^\prime_{\rm{peak}}$,  where $E^\prime_{\gamma, \rm{prompt}}$ is the total energy in the prompt emission in the comoving frame and $E^\prime_{\rm{peak}}$ is the peak energy of the spectrum in the comoving frame, representing an average of $h\nu^\prime$. Since the number of photons is a Lorentz invariance, it also equals $E_{\gamma, \rm{prompt}}/(E_{\rm{peak}}(1+z))=E_{\rm{iso}}f_b/(E_{\rm{peak}}(1+z))$. We therefore have the relation:
\begin{equation}
(1-\eta)N_{\gamma,0}\approx E_{\rm{iso}}f_b/(E_{\rm{peak}}(1+z)). 
\label{eq:20}
\end{equation}
Taking equation (\ref{eq:20}) into the expression of $\tau_{\gamma\gamma}$ into equation (\ref{eq:17}), we find that:
\begin{equation}
\tau_{\gamma\gamma}=\sigma_{\gamma\gamma}\frac{\eta}{1-\eta}\frac{E_{\rm{iso}}}{E_{\rm{peak}}(1+z)}R^{-2}_{\rm{prod}}>1.
\end{equation}
The above inequality can be used to set a lower limit of the $E_{\rm{iso}}$ of GRB which can produce \pair:
\begin{eqnarray}
E_{\rm{iso}}&>&E_{\rm{peak}}(1+z)\frac{1-\eta}{\eta}\sigma_{\gamma\gamma}^{-1}R^2_{\rm{prod}}\nonumber\\
&\gtrsim&1.39\,E_{\rm{peak}}(1+z)\frac{1-\eta}{\eta}\sigma_{\rm T}^{-1}R^2_{\rm{prod}},
\end{eqnarray}
as the maximum value of $\sigma_{\gamma\gamma}$ is about 0.72 $\sigma_{\rm T}$.
%${\rm Max}\left(\sigma_{\gamma\gamma}\right)\simeq0.72~\sigma_{\rm T}$.}
Now, suppose the energy in the photons which participated the pair production is less than $\sim10\%$ of those which did not (which can be demonstrated via detailed simulation). The former are photons with $h\nu^\prime\gtrsim$\,MeV (so that $h\nu^\prime_1+h\nu^\prime_2>2m_ec^2$), and the latter are those with $h\nu^\prime\sim$\,keV ($E^\prime_{\rm{peak}}\sim$\,keV). As a result, the number ratio $\eta/(1-\eta)$ between them, should be less than 10$^{-4}$. We take $\eta/(1-\eta)\lesssim10^{-4}$ into the above lower limit of $E_{\rm{iso}}$ to obtain: 
\begin{equation}
E_{\rm{iso}}\gtrsim3.3\times10^{53}E_{\rm{peak},100}(1+z)R^2_{\rm{prod},16}\,\text{erg},
\label{eq:limit}
\end{equation}
where $R_{\rm{prod},16}$ is $R_{\rm{prod}}$ in unit of $10^{16}$\,cm, and $E_{\rm{peak},100}$ is the observed $E_{\rm{peak}}$ in unit of 100 keV. Equation (\ref{eq:limit}) gives  a limit in the parameters space of the Amati relation ($E_{\rm{iso}}$-$E_{\rm{peak}}(1+z)$), where \pair production in GRB prompt emission is only possible in those sources above this limit (see figure \ref{fig:limit}). As we can see from figure \ref{fig:limit}, there are a few candidates lie beyond the limit line when $R_{\rm{prod},16}=1$, and there are tens of candidates when $R_{\rm{prod},16}=0.3$. We list these candidates in tables \ref{tab:1} and \ref{tab:2}. 

It should be noted that, passing the criterion (equation \ref{eq:limit}) alone  does not guarantee annihilation line be observed in those GRBs, for there are more requirements such as fast cooling, efficient annihilation and optically thin conditions for the line photons to emerge. {\revise For instance, the fast cooling of \pair pairs is closely related to the luminosity of the GRB.} 

{\revise Furthermore, the detectability of the emission line depends not only on whether the line photons can be produced and escape but also on more factors, such as the relative strength of the spectral line compared to the continuum prompt emission, the flux or fluence of the line emission, detectors’ energy and time resolution, as well as the background level during observation. A detailed and more quantitative study would be necessary to provide a reliable prediction on the detectability. In reality, it would be more convenient to directly conduct line searches in the archival data of the selected burst candidates listed in Table \ref{tab:1}}. 
%去掉 \textit{etc}，避免与such as 重复

%in terms of their fluence, 

%Our result indicate that, to trigger the \pair production in prompt emission, the GRB should have an very high $E_{\rm{iso}}$ larger than $\sim10^{54}$\,erg. %Therefore, the GRB221009A, whose $E_{\rm{iso}}=1/5\times10^{55}$\,erg, may remain the only GRB emitting \pair annihilation line in its prompt emission period. 
\section{Conclusion and discussion}
\subsection{Summaries on the conclusions and physical scenario of pair production and annihilation}
\begin{figure}
\centering
\includegraphics[width=12 cm]{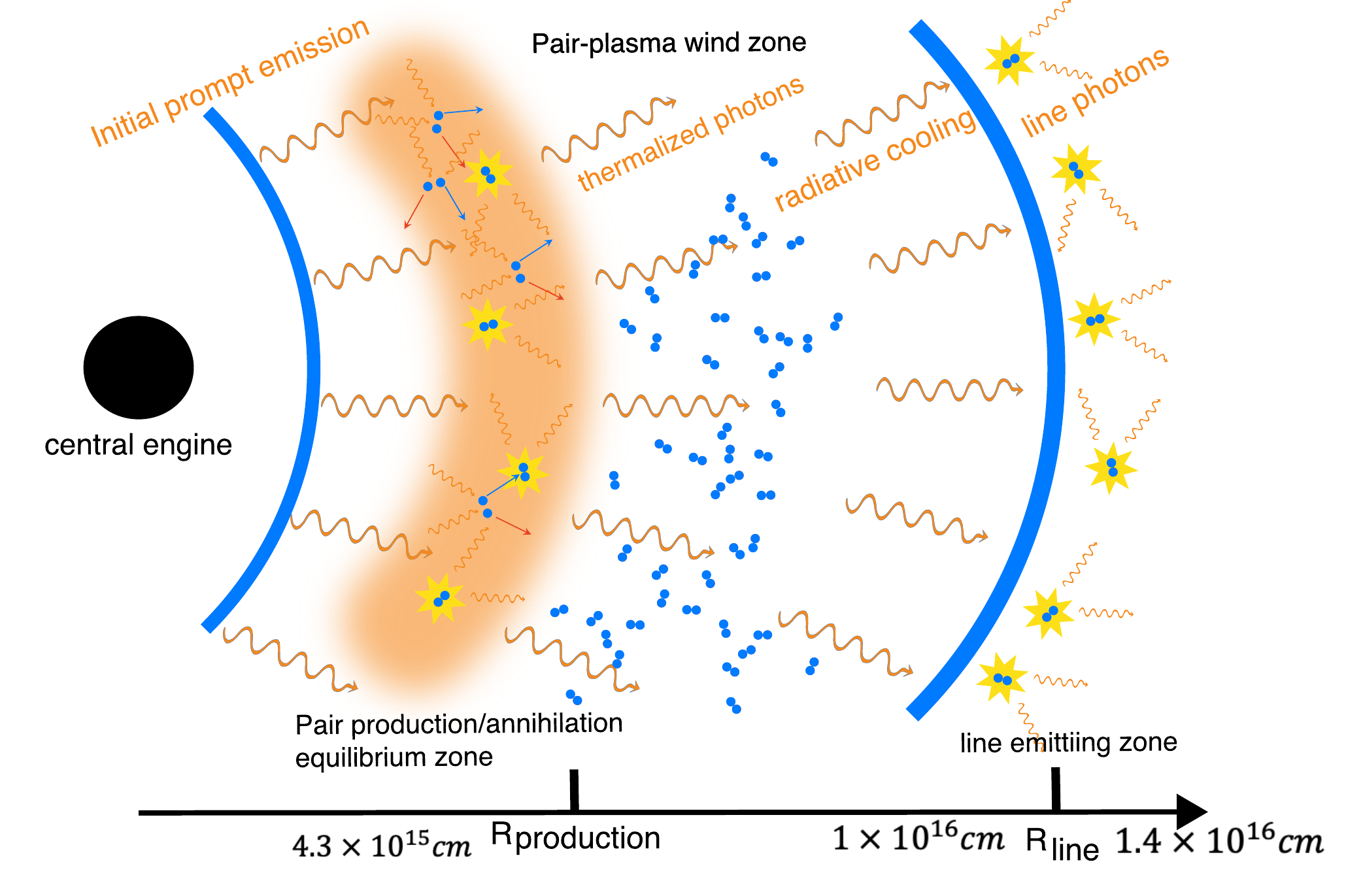}
\caption{The schematic of the physical processes of the production of \pair annihilation line emission.}
\label{fig:2}
\end{figure}

In the above section, we took arguments on the optical depths of several physical process, and drew conclusions based on them:
\begin{itemize}
\item{The line emission region should be optically thin, otherwise the annihilation line will not emerge $\longrightarrow$ $R_{\rm{line}}>1.4\times10^{16}$\,cm;}\\
\item{The annihilation optical depth between \pair should be larger than unity, in order to effectively produce the line; besides, \pair cannot be efficiently cooled down to $\beta^\prime_e<0.01$ $\longrightarrow$ $R_{\rm{line}}<4.3\times10^{16}$\,cm;}\\
\item{In the location $R_{\rm{prod}}$ where \pair are produced, the annihilation optical depth between \pair should be less than unity, otherwise the pairs will not be preserved to $R_{\rm{line}}$ $\longrightarrow$ $R_{\rm{prod}}>4.3\times10^{15}$\,cm;}\\
\item{The optical depth of $\gamma\gamma\rightarrow e^-/e^+$ needs to be larger than unity in the prompt emission region, in order to produce \pair $\longrightarrow$ A GRB needs to be bright enough that $E_{\rm{iso}}\gtrsim3.5\times10^{53}E_{\rm{peak},100}(1+z)R^2_{\rm{prod},16}\,\text{erg}$ to have \pair annihilation production. Candidate GRBs that could have \pair produced in their prompt emission are listed in tables \ref{tab:1} and \ref{tab:2}. }
\end{itemize}
The complete scenario of the \pair production and emission of the annihilation line is summarized as follows:
\begin{enumerate}
\item{The prompt emission photons are radiated at a radius $\sim10^{15}$ cm. In these photons, a fraction of them satisfy the condition that $h\nu^\prime_1+h\nu^\prime_2\ge 2m_ec^2$ in the jet comoving frame. }
\item{The number of those high energy portion of photons are so large that they can efficiently produce \pair. \pair and the high energy photons are in an equilibrium state of the pair production/annihilation process. Therefore, the number density of \pair tracks that of the high energy photons. % {\red During this stage, there is no significant increase in the luminosity of the prompt emission. Actually, in the lightcurve of GRB 221009A, a slow-rising part can be identified in the first bump, just prior to its fast-rising episode; see panel (d) of figure 1 in Zhang24 for further details.}
}
\item{The equilibrium $\gamma\gamma\leftrightarrow e^-e^+$ process thermalize those high energy portion of photons, and the number density of \pair decreases to unity, when the annihilation is no longer efficient.} %{\red Accordingly, a thermalized component with temperature $T_{\rm dec}$ can be expected to be produced in the spectra just after decoupling.}}
\item{\pair propagates from this distance $R_{\rm{prod}}$ to a larger distance $R_{\rm{line}}$. During the course, the \pair is cooled down due to synchrotron or inverse Compton radiation. }
%{\red This stage roughly corresponds to the fast-rising episode of the prompt emission, unlikely occurring after the main burst. In GRB 221009A, the electrons and positrons are more likely cooled through the synchrotron radiation mechanism. During this stage, the low-energy spectrum index $\alpha$, the high-energy spectrum index $\beta$, and the peak energy $E_{\rm peak}$ are \citep{An23}: $-\alpha\sim0.8-1$, $\beta\sim-2.5$, and $E_{\rm peak}\sim0.5$ MeV, respectively. These values are basically in line with the predicted ones by the synchrotron radiation mechanism \citep{zhang19} in the fast cooling regime \citep[e.g.,][]{Uhm:2013gwa}. The fast cooling, which is required by the narrow width of the MeV emission line, also indicates the appearance of a strong magnetic field in relevant regions. Accordingly, the thermal component mentioned above may be reshaped by the magnetic field, and thus, it can manifest as a low-energy excess or bump in the corresponding observed spectrum. }}
\item{\pair is cooled down to a critical $\beta^\prime_e$ in the comoving frame, so that the annihilation process become efficient again.} %{\red As indicated by the narrow line width, we have $\beta^\prime_e\sim0.1$ in GRB 221009A.}}
\item{\pair annihilate fast, and the emitted photons can freely escape to infinity due to the optically thin environment.}

\end{enumerate}
The above scenario is depicted in figure \ref{fig:2}. 

%In such a scenario, there is a time lag $\Delta T\sim R_{\rm{prod}}/c\sim10^{5}$\,s between the peak prompt emission time and the line occurrence time in the burst central frame.
%That $\Delta T$ corresponds to a time lag in the observer frame as $\Delta t=\Delta T/\Gamma^2\sim10$\,s, which is in agreement with the observation. 

It takes $\Delta T_{\rm dyn}\approx(R_{\rm{line}}-R_{\rm{prod}})/(\beta_ec)$ for the pairs to propagate from $R_{\rm{prod}}$ to their annihilation site $R_{\rm{line}}$. This corresponds to a observed time delay $\delta t_{\rm{dyn}}=(1+z)\Delta T_{\rm{dyn}}/\Gamma^2\sim10$ s, which is in agreement with the observation that, the time lag between the prompt emission peak and the first occurrence of the line is $\sim10$ s. 
%{\red Let us define a comoving dynamical timescale as
% $\Delta t^{\prime}_{\rm dyn}\sim(R_{\rm{line}}-R_{\rm{prod}})/(\Gamma c)$, where $\Gamma$ is the jet bulk Lorentz factor.
% %$\Delta t^{\prime}_{\rm dyn}\sim\frac{R_{\rm{line}}R_{\rm{prod}}}{\Gamma c}$.
% In such a scenario, it is actually a time lag between the peak prompt emission time and the line occurrence time in the comoving frame. In the observer frame, it corresponds to a time lag of $\Delta t_{\rm dyn}=\Delta t^{\prime}_{\rm dyn}/\Gamma\sim10$ s, which is in agreement with the observation.}

%{\red In such a scenario, the dynamical timescale is $t^{\prime}_{\rm dyn}\sim R_{\rm{prod}}/(\Gamma c)\sim10^{5}/\Gamma\,{\rm s}$ in the comoving frame. In the observer frame, it corresponds to a time lag between the peak prompt emission time and the line occurrence time as $t_{\rm dyn}=t^{\prime}_{\rm dyn}/\Gamma\sim10$\,s, which is in agreement with the observation.}
\subsection{Prediction on the relic thermal radiation from pair production}
As mentioned above, a major prediction of this scenario is that there should be a thermal component emerging between the epochs of the peak prompt emission and the appearance of line emission. The energy density of the thermalised photons in the comoving frame is equal to that of the photons that participated in \pair production: 
\begin{equation}
\epsilon^\prime\sim n^\prime_\gamma h\nu^\prime,
\end{equation}
where $h\nu^\prime$ denotes the averaged energy of pair producing photons. According to equation (\ref{eq:21}), which relates $n^\prime_\gamma$ with $n^\prime_{\pm}$, the energy density can be further expressed as:
\begin{equation}
\epsilon^\prime\sim n^\prime_{\pm}{\gamma^\prime_e}^{3/2}m_ec^2,
\end{equation}
where, again, $\gamma^\prime_e\equiv h\nu^\prime/m_ec^2$. On the other hand, the corresponding black body radiation temperature $T^\prime$ is related with the energy density $\epsilon^\prime$ as:
\begin{equation}
\epsilon^\prime=a{T^\prime}^4,
\end{equation}
where $a$ is related to the Stefan-Boltzmann constant $\sigma$ as: $a=4\sigma/c$. As a result, we can infer that the temperature of the thermal component in the comoving frame is that:
\begin{equation}
T^\prime=\large\left(n^\prime_{\pm}{\gamma^\prime_e}^{3/2}m_ec^2/a\large\right)^{1/4}.
\label{eq:29}
\end{equation}
The peak luminosity of this thermal component is then:
\begin{eqnarray}
L_{\rm{th,peak}}&=&\delta\Omega\, R^2_{\rm{prod}}\sigma{T^\prime}^4\Gamma^2\nonumber=(1/\Gamma)^2R^2_{\rm{prod}}\sigma{T^\prime}^4\Gamma^2\nonumber\\
&=&\frac{1}{4}R^2_{\rm{prod}}cn^\prime_{\pm}{\gamma^\prime_e}^{3/2}m_ec^2\nonumber\\
&=&\frac{1}{4}R^2_{\rm{prod}}c{\gamma^\prime_e}^{3/2}m_ec^2\frac{N_{\pm}}{V^\prime},
\end{eqnarray}
where $\delta\Omega=1/\Gamma^2$ is the solid angle over which the thermal radiation body is observable due to Doppler beaming effect. Taking the expressions of $N_{\pm}$ and $V^\prime$ (substituting $R_{\rm{line}}$ with $R_{\rm{prod}}$) in equations (\ref{eq:3}) and (\ref{eq:4}), one further expresses $L_{\rm{th,peak}}$ as:
\begin{equation}
L_{\rm{th,peak}}=\frac{c{\gamma^\prime_e}^{3/2}m_ec^2(1+z)}{4R^\prime_{\rm{prod}}}\frac{\mathcal{F}_0}{\mathcal{E}_{{\rm{line},0}}}\ln\frac{t_f-t_0}{t_i-t_0}D^2_{\rm{L}}.
\label{eq:31}
\end{equation}
As $\delta R^{\prime}_{\rm{prod}}$ is the width of the pair production zone in the comoving frame, it is related to that in the rest frame $\delta R_{\rm{prod}}$ with $\delta R^{\prime}_{\rm{prod}}=\Gamma\delta R_{\rm{prod}}$. Taking numerical values into equations (\ref{eq:29}) and (\ref{eq:31}), one can get that: $T^\prime=1.38\times10^4\gamma^{\prime\,3/8}_{e,100}\,\text{K}$, where $\gamma^\prime_{e,100}$ is $\gamma^\prime_e$ in unit of 100. Therefore, the maximum Doppler boosted temperature of the thermal component is $T=(1+z)\,\Gamma T^\prime\sim0.1-1$~keV, 
whose radiation is in the soft X-ray band to be distinguished from the photospheric thermal emission of a GRB at much higher energies. The peak luminosity is $L_{\rm{th,peak}}=2\times10^{44}\,\gamma^{\prime\,3/2}_{e,100}\,\delta R_{\rm{prod,16}}~\text{erg/s}$. Correspondingly, the peak flux is $F_{\rm{th,peak}}=3.0\times10^{-12}\,\gamma^{\prime\,3/2}_{e,100}\,\delta R_{\rm{prod,16}}~{\rm erg/cm^2/s}$. %It is already much higher than the accuracy of $\sim10^{-13}\,{\rm erg/s/cm^2}$ achieved during the soft X-ray afterglow observations of GRB 060614. This means that the thermal component can be detected in the prompt emission stage with current telescopes, like ... , as well as by future telescopes, such as ... . 
As in Zhang24, the authors explain the apparent line centre energy evolution as a result of the high-latitude effect. Similar effect will also change the thermal radiation, making the apparent temperature of the thermal component evolve with time $kT\propto(t-t_0)^{-1}$, and its flux evolves with $F_{\rm{th}}(t)\propto(t-t_0)^{-3}$ \citep{2000ApJ...541L..51K,2015MNRAS.450.3549S}. Therefore, we expect the thermal radiation to sweep through $\sim$ keV to $\sim$0.1 keV in soft X-ray band in tens of seconds, with its flux decreasing from $\sim10^{-12}$ to $\sim10^{-15}$ ${\rm erg/cm^2/s}$. The flux is roughly comparable to that of $\sim10^{-13}~{\rm erg/cm^2/s}$ (in 100 s at 0.5-2 keV) achieved by near future telescopes like Athena/WFI \citep{Piro2022}. Nevertheless, such a signal is short duration transient, thus its detection is far below the sensitivity of the current all-sky soft X-ray monitor such as the Wide-fiend X-ray Telescope (WXT) onboard the Einstein Probe \citep{2022hxga.book...86Y} or any all-sky soft X-ray monitor to be deployed in a foreseeable future.

%%$T=(1+z)\,\Gamma T^\prime=7\times10^6$~K, ultraviolet
\section{Data Availability}
No new data were generated or analysed in support of this research.
\section*{Acknowledgments}
YSX would like to thank useful discussion over this topic with Profs. Bing Zhang and Zhuo Li. This work is supported by the fund from the Chinese Academy of Sciences (grant Nos. E329A3M1 and E3545KU2)

\appendix

\section{The relative velocity between two relativistic particles}
\label{sec:app}
The relative velocity between two particles (with velocities $\beta_1$, $\beta_2$) is \citep{LandauLifshitz2002}: 
\begin{equation}
\beta^2_{\rm{rel}}=\frac{\beta_1^2+\beta_2^2-2\beta_1\beta_2\cos\theta-\beta^2_1\beta^2_2\sin^2\theta}{(1-\beta_1\beta_2\cos\theta)^2}.
\end{equation}
Integration of the above equation over the solid angle gives the directional averaged relative velocity: 
\begin{equation}
\bar{\beta^2_{\rm{rel}}}=\frac{1}{2}\int_0^\pi\beta^2_{\rm{rel}}\sin\theta d\theta=\frac{\beta_1^2+\beta_2^2-2\beta_1^2\beta_2^2}{1-\beta_1^2\beta_2^2}. 
\end{equation}
Therefore, if all particles have the same velocity $\beta^\prime_e$, one finds: 
\begin{equation}
\bar{\beta}^2_{\rm{rel}}=2\frac{\beta^{\prime 2}_e}{1+\beta^{2 \prime}_e}.
\end{equation} \\

\newpage
\section{Candidates that could have \pair produced in their prompt emission}
In this appendix section, we give tables \ref{tab:1} and \ref{tab:2} of GRB candidates that could have \pair production in their prompt emission stages, according to criterion equation (\ref{eq:limit}). 

\begin{table}
\centering
\begin{minipage}{\textwidth}
\centering
%\captionsetup{justification=centering, labelfont=bf}
\caption{GRB candidates which could have \pair production in their prompt emission stages, according to criterion equation (\ref{eq:limit}), assuming $R_{\rm{prod, 16}}=1$}
\begin{tabular*}{\textwidth}{@{\extracolsep{\fill}} lcccccc}
\toprule
GRB & 
\( z \) & 
\( T_{90} \) (s) & 
\( E_{p,i} \) (keV) & 
\( E_{\rm iso} \) (\(10^{52}\) erg) & 
Detector \\
\midrule
110918A & 0.98 &   19.60 &  667.0  &  270.50 & FER \\
130907A & 1.23 &  360.00 &  882.0 &  314.00 & FER \\
160625B & 1.40 &  454.67 & 1374.0 &  510.10 & FER \\
180914B & 1.09 &  272.48 &  977.0 &  370.00 & KW  \\
190530A & 0.93 &   18.43 & 1745.0 &  605.00 & FER \\
210619B & 1.93 &   54.79 &  598.0 &  344.00 & FER \\
221009A & 0.15 &  600.00 & 1436.0 & 1500.00 & FER/KW/HXMT/GECAM \\
\bottomrule
\end{tabular*}
\label{tab:1}
%\vspace{0.5cm}
\begin{flushleft}
\textbf{Notes:} Column (1): GRB name. Column (2): redshift. Column (3): value of \(T_{90}\). Column (4): isotropic \(\gamma\)-ray energy in rest-frame between \( 1 - 10^{4}\) keV. Column (5): Peak energy in rest frame (intrinsic). Column (6): the experiment, used for the measurement of \(T_{90}\), \(E_{p}\), and \(E_{\rm iso}\) values (FER = Fermi, KW = Konus-Wind, SWI = Swift, HXMT = Hard X-ray Modulation Telescope, GECAM = Gravitational wave high-energy Electromagnetic Counterpart All-sky Monitor). Data taken from \cite{An23}.
\end{flushleft}
\end{minipage}

\end{table}

\begin{table}
\centering
\caption{GRB candidates which could have \pair production in their prompt emission stages, according to criterion equation (\ref{eq:limit}), assuming $R_{\rm{prod},16}=0.3$}
\begin{tabular*}{\textwidth}{@{\extracolsep{\fill}} lcccccc}
\toprule
GRB & 
\( z\) & 
\( T_{90} \) (s) & 
\( E_{p,i} \) (keV) & 
\( E_{\rm iso} \) (\(10^{52}\) erg) & 
Detector \\
\midrule
970828 & 0.96 & 146.59 &  586.0 &   30.38 & GRO \\
971214 & 3.42 &   6.00 &  685.0 &   22.06 & SAX \\
990123 & 1.60 &  61.00 & 1724.0 &  242.38 & SAX/GRO/KW \\
990506 & 1.30 & 129.00 &  677.0 &   98.13 & SAX/GRO/KW \\
990510 & 1.62 &  57.00 &  423.0 &   17.99 & SAX \\
990705 & 0.84 &  32.00 &  459.0 &   18.70 & SAX/KW \\
991208 & 0.71 &  63.10 &  313.0 &   22.97 & KW \\
991216 & 1.02 &  45.00 &  648.0 &   69.79 & SAX/GRO/KW \\
000131    & 4.50 &  96.30 &  987.0 &  181.48 & GRO \\
000418    & 1.12 &   2.00 &  284.0 &    9.51 & GRO \\
000911    & 1.06 &  23.30 & 1856.0 &   69.86 & GRO \\
000926    & 2.04 &  54.70 &  310.0 &   27.98 & GRO \\
010222  & 1.48 &  74.00 &  766.0 &   85.57 & GRO \\
011211  & 2.14 &  51.00 &  186.0 &    5.71 & GRO \\
020124  & 3.19 &  51.17 &  448.0 &   27.02 & GRO \\
020405  & 0.69 &  40.00 &  354.0 &   10.64 & GRO \\
020813  & 1.26 &  87.34 &  590.0 &   68.35 & GRO \\
030226  & 1.99 &  76.23 &  289.0 &   12.94 & GRO \\
030328  & 1.52 & 138.27 &  328.0 &   39.42 & GRO \\
030528  & 0.78 &  62.80 &   57.0 &    2.22 & GRO \\
040912  & 1.56 &   9.21 &   44.0 &    1.36 & GRO \\
041006  & 0.72 &  22.08 &   98.0 &    3.11 & GRO \\
050401  & 2.89 &  33.30 &  467.0 &   36.39 & GRO \\
050603  & 2.82 &  11.08 & 1333.0 &   64.03 & GRO \\
050814  & 5.30 & 144.00 &  339.0 &   11.20 & GRO \\
050820  & 2.61 &  26.00 & 1325.0 &  102.89 & GRO \\
050904  & 6.29 & 181.70 & 3178.0 &  127.35 & GRO \\
051008  & 2.77 &  64.00 & 2074.0 &   83.40 & GRO \\
051022  & 0.81 & 178.00 &  754.0 &   56.04 & GRO \\
060124  & 2.29 & 658.20 &  784.0 &   43.85 & GRO \\
060210  & 3.91 & 242.18 &  575.0 &   41.53 & GRO \\
060510B & 4.90 & 262.95 &  575.0 &   36.70 & SWI \\
060714  & 2.71 & 116.04 &  234.0 &   13.40 & SWI \\
060814  & 1.92 & 144.95 &  751.0 &   56.71 & SWI \\
060906  & 3.69 &  44.58 &  209.0 &   14.90 & SWI \\
060927  & 5.46 &  22.42 &  475.0 &   14.49 & SWI \\
061007  & 1.26 &  75.74 &  890.0 &   89.96 & SWI \\
061222A & 2.09 &  96.00 &  874.0 &   30.04 & SWI \\
061222B & 3.36 &  37.25 &  200.0 &   10.30 & SWI \\
070125  & 1.55 &  70.00 &  934.0 &   84.62 & KW \\
070328  & 2.06 &  72.12 & 1182.0 &   77.70 & KW \\
080319B & 0.94& 124.86 & 1261.0 &  117.87 & SWI/KW \\
\bottomrule
\end{tabular*}
\label{tab:2}
\end{table}

\begin{flushleft}
\textbf{Notes:} Excludes GRBs already satisfying constraints set by \( R_{\rm{prod},16} = 1 \).  Column (1): GRB name. Column (2): redshift. Column (3): value of \(T_{90}\). Column (4): isotropic \(\gamma\)-ray energy in rest-frame between \( 1 - 10^{4}\) keV. Column (5): Peak energy in rest frame (intrinsic). Column (6): the experiment, used for the measurement of \(T_{90}\), \(E_{p}\), and \(E_{\rm iso}\) values (GRO = CGRO/BATSE, FER = Fermi, KW = Konus-Wind, SWI = Swift, HXMT = Hard X-ray Modulation Telescope, GECAM = Gravitational wave high-energy Electromagnetic Counterpart All-sky Monitor). Data taken from \cite{An23}
\end{flushleft}

%\bibliography{sample631}{}
%\bibliographystyle{}

\end{document}